 \documentclass[twocolumn]{aastex6}
\usepackage{amsmath}
\usepackage{natbib}
\begin{document}
\title{Pre-flare coronal jet and evolutionary phases of a solar eruptive prominence associated with M1.8 flare: SDO and RHESSI observations}

\author{Bhuwan Joshi}
\affil{Udaipur Solar Observatory, Physical Research Laboratory, Udaipur 313001, India}

\author{Upendra Kushwaha}
\affil{Udaipur Solar Observatory, Physical Research Laboratory, Udaipur 313001, India}

\author{Astrid M. Veronig}
\affil{Kanzelh\"ohe Observatory/Institute of Physics, University of Graz, Universit$\ddot{a}$tsplatz 5, A-8010 Graz, Austria}

\author{K. -S. Cho}
\affil{Korea Astronomy and Space Science Institute, Daejeon 305-348, Korea}

\begin{abstract}
We investigate triggering, activation, and ejection of a solar eruptive prominence that occurred in a multi-polar flux system of active region NOAA 11548 on 2012 August 18 by analyzing data from AIA on board SDO, RHESSI, and EUVI/SECCHI on board STEREO. Prior to the prominence activation, we observed striking coronal activities in the form of a blowout jet which is associated with rapid eruption of a cool flux rope. Further, the jet-associated flux rope eruption underwent splitting and rotation during its outward expansion. These coronal activities are followed by the prominence activation during which it slowly rises with a speed of $ \sim $12~km~s$^{-1}$ while the region below the prominence emits gradually varying EUV and thermal X-ray emissions. From these observations, we propose that the prominence eruption is a complex, multi-step phenomenon in which a combination of internal ({\it tether-cutting} reconnection) and external (i.e., pre-eruption coronal activities) processes are involved.  The prominence underwent catastrophic loss of equilibrium with the onset of the impulsive phase of an M1.8 flare suggesting large-scale energy release by coronal magnetic reconnection. We obtained signatures of particle acceleration in the form of power law spectra with hard electron spectral index ($\delta$ $\sim$3) and strong HXR footpoint sources. During the impulsive phase, a hot EUV plasmoid was observed below the apex of the erupting prominence that ejected in the direction of the prominence with a speed of $ \sim $177~km~s$^{-1}$. The temporal, spatial and kinematic correlations between the erupting prominence and the plasmoid imply that the magnetic reconnection supported the fast ejection of prominence in the lower corona.

\end{abstract}


\section{Introduction}
\label{sec:intro}
Exploration of source region characteristics of prominence eruptions is of utmost importance in view of understanding the triggering mechanisms for solar eruptions and to predict their possible space weather consequences. With the availability of multi-wavelength data, especially from space-based platforms, it has become clear that different manifestations of solar eruptions (i.e., eruptive prominences, flares, and coronal mass ejections) are closely related with the process of disruption of coronal magnetic fields. It is widely accepted that magnetic reconnection is the most fundamental process responsible for the changes in the topology of coronal magnetic fields, as well as the rapid conversion of stored magnetic energy into thermal and kinetic energy of plasma and particles during solar eruptive events \citep[e.g., see review by][]{Priest2002, Fletcher2011, Wiegelmann2014}.

Before the activation of a prominence (or equivalently filament when viewed on the solar disk), its relatively cool, dense material and associated magnetic structure remain suspended in the hotter corona by magnetic fields. When these structures erupt, both prominence material and magnetic fields are expelled together leading to radiative signatures at multi-wavelengths in the source region and a coronal mass ejection (CME) beyond the solar atmosphere. Filaments are commonly classified into two categories: active region and quiescent \citep{Tandberg-Hanssen}. Active region filaments are low-lying and rapidly evolving structures, forming in the newly emerged magnetic fields of an active region. They do not protrude much over the solar limb. On the other hand, quiescent prominences are generally associated with the decaying phase of an active region and are long lived. They tend to be located high in the corona. The eruptions of active region prominence are frequently associated with solar flares. In case when the amount of energy release is higher, the source region exhibits evidence of high energy processes in the form of X-ray and MW sources and emission in hot EUV channels during flare's impulsive phase. Notably, recent studies reveal that there is a close physical connection between the impulsive phase of solar flares and the main acceleration phase of CME \citep{Zhang2001,Maricic2007, Temmer2008}. 

The relationship between filament rise and associated radiative signatures (i.e., flare or pre-flare activity) has been the subject of several recent studies \citep{LiuW2009,Joshi2011,Sterling2011,Joshi2013,Kushwaha2015}. These studies reveal that, in general, the active region filament eruption undergoes two phases of evolution: a slow rising activation phase which is dominated by thermal emission (preheating phase), and eruptive phase during which impulsive flare emission is observed. With the advent of Atmospheric Imaging Assembly (AIA) on board Solar Dynamics Observatory (SDO), the case studies of prominence eruptions have become much more promising as the source region of eruption can be simultaneously monitored with multi-channel EUV/UV images having unprecedented spatial and temporal resolutions. If available, it is quite useful to combine EUV observations with multi-band X-ray time-profiles, images, and spectra from the Reuven Ramaty High Energy Solar Spectroscopic Imager (RHESSI). The comparisons of location, timing, and strength of high energy emissions with respect to the dynamical evolution of the prominence provide critical clues to understand the characteristics of the underlying energy release phenomena, such as, expected site of magnetic reconnection, particle acceleration, heating, etc. \citep{LiuR2009,Alexander2006,Chifor2006,Vema2012,Joshi2013,Awasthi2014}.

A very important objective toward the investigations of the source regions characteristics of solar eruptions is to address their triggering mechanisms {\citep{Manoharan2003,Joshi2007,Sterling2011}. Several models and magnetic configurations have been proposed to exploit the possible mechanisms toward the onset of eruptive flares \citep[see reviews by][]{Lin2002,Aulanier2014}. The {\it flux cancellation model} exhibits that the dissipation of magnetic flux at the surface reduces the magnetic tension force of the overlying field that confines a flux rope so that the upward magnetic pressure dominates at some point of the evolution leading to the eruption of the flux rope \citep{Linker2003}. In the {\it tether-cutting model}, it has been proposed that the eruption in a bipolar system can be triggered by magnetic reconnection within the sheared core field resulting the expulsion of overlying sigmoid structures \citep{Moore1992,Moore2001}. The {\it breakout} model} explains the ejection of flux rope in the multi-polar system \citep{Antiochos1999, Karpen2012}. In this model, the early reconnection occurs in a current sheet formed at the interface between the inner and overlying flux domains. This reconnection reduces the confining tension of the overlying fields resulting the inner domain to expand. When the inner sheared fields have expanded outward sufficiently, a vertical current sheet forms in the low corona. The large-scale reconnection in this vertical current sheet is responsible for the flare emission. In observations, the identification of the suitable triggering mechanism for a solar eruption needs a thorough examination of the activities near the activation site and surrounding regions well before the start of the filament's activation. These activities may not always be associated with noticeable radiative signatures but can represent important features related to the dynamical evolution of the pre-eruption corona. 

In this study, we present the multi-channel EUV and X-ray observations of dynamical activities observed in the active region NOAA 11548 on 2012 August 18 over an hour. The focus is on the evolutionary stages of an active region prominence which undergoes a very interesting dynamical evolution together with associated radiative signature in the form of pre-flare heating, precursor emission and an eruptive M1.8 flare. The erupted filament leads to a partial-halo, fast CME with a linear speed of 834~km~s$^{-1}$. We have further undertaken a detailed investigation on the temporal and spatial evolution of the X-ray emission during the early expansion and subsequent eruption of the prominence. We provide a note on observing instruments in section~\ref{sec:obs}. In section~\ref{sec:results}, we provide the details of our analysis and describe the observational results. We discuss and interpret our results in  	section \ref{discuss}. The conclusions drawn from this study are given in the final section of the paper. 

\section{Observations}
\label{sec:obs}
Our analysis is mainly based on observations taken by AIA \citep{Lemen2012} on board SDO and RHESSI \citep{Lin2002}. AIA records full-disk images at 12 s cadence in seven different EUV filters (94, 131, 171, 193, 211, 304, and 335~\AA), at 24 s cadence in two UV filters (1600 and 1700~\AA), and at 3600 s cadence in white light filter (4500~\AA). In this paper, we present AIA images taken in 94, 131, 171, and 193~\AA~band passes with a spatial sampling of 0$\arcsec $.6 pixel$ ^{-1} $. During the flare activity, AIA takes images with different exposure times ranging from 0.1 to 2.9 s. To compensate for this, each AIA image was normalized to its respective exposure time. We have also analyzed EUV images at 195 \AA~obtained from the Extreme Ultraviolet Imager (EUVI; \citealt{Wuelser2004}) which is part of the Sun-Earth-Connection Coronal and Heliospheric Investigations (SECCHI; \citealt{Howard2007}) instrument suite on board NASA STEREO mission. EUVI observes full Sun (upto 1.7 R$ _{\odot} $) with a spatial resolution of 1.6$ ^{\prime\prime} $ pixel$ ^{-1} $ and a temporal cadence of 5 minutes. The EUVI images analyzed here correspond to STEREO-B observations which had an angular separation of -115$ ^{\circ} $ with respect to Earth (in heliocentric coordinates) on the day of event ({http:$//stereo-ssc.nascom.nasa.gov/cgi-bin/make\_where\_gif$}).     

RHESSI observations have revealed significant activities in NOAA 11548 during its continuous observations from 02:45 UT to 03:45 UT on 2012 August 18. RHESSI observes the full Sun with an unprecedented combination of spatial resolution (as fine as $ \sim $2$ {\arcsec} $.3) and energy resolution (1--5~keV) in the energy range of 3~keV to 17~MeV. For image reconstruction, depending upon the requirements, we have used CLEAN and PIXON algorithms. The images are reconstructed with the natural weighing scheme using front detector segments 3--8
(excluding 7).
               
\section{Data analysis and results}
\label{sec:results}

\subsection{Phases of prominence eruption}
\label{sec:prominence}

\begin{figure*}
\epsscale{1.0}
\plotone{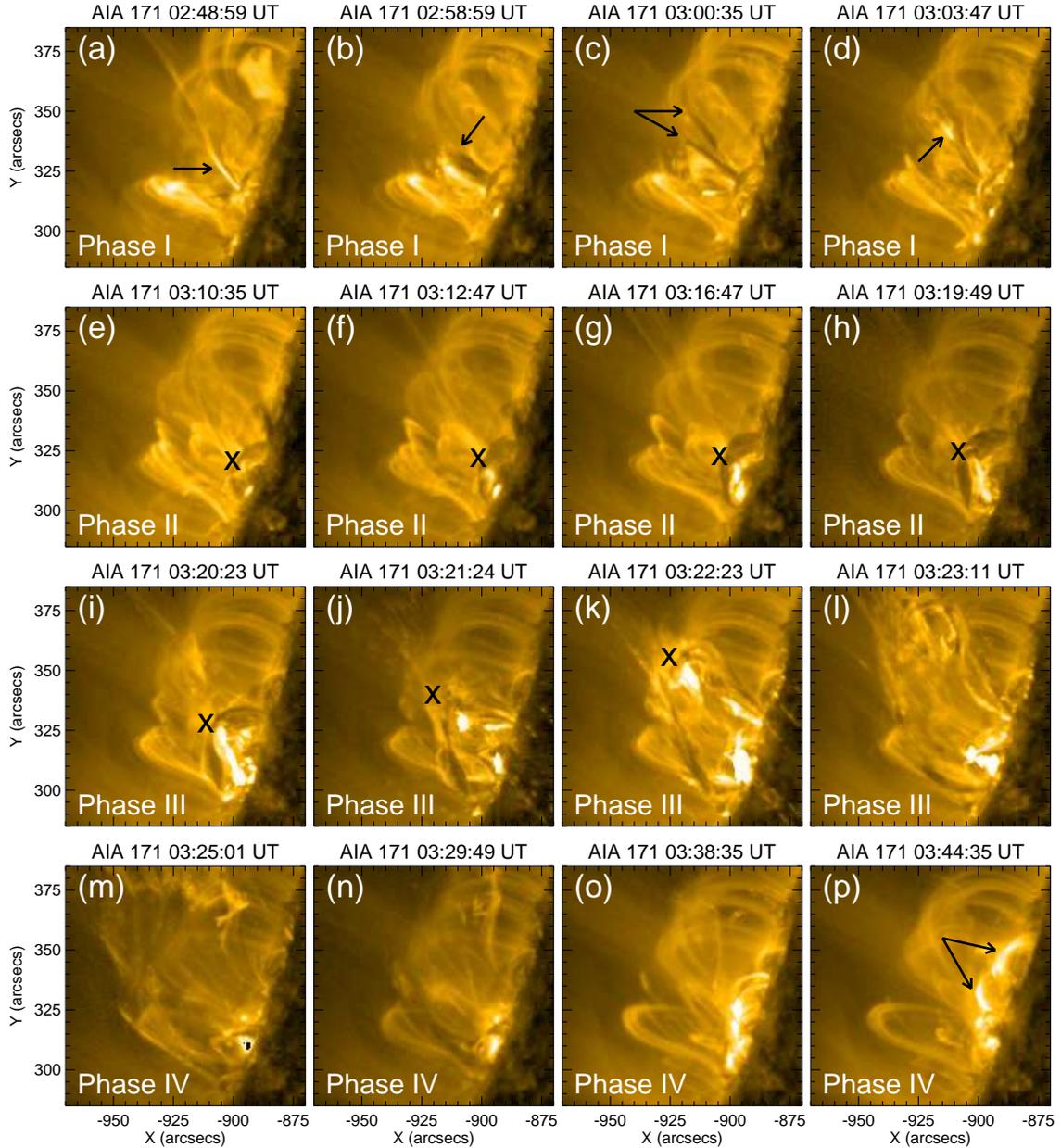}
\caption{Series of AIA~171 \AA~images showing the overall eruption scenario which is divided in four phases (see Table~\ref{tab:phases}). Several important features are identified during the various phases which are annotated in this figure. Phase I: Arrows indicate bright spiky structure (panel (a)), fast rise and evolution of a flux rope (panels (b)-(d)); Phase II and III: The prominence apex (marked by cross ($\times$)) during its slow and fast rise phases (panels (e)-(k)); Phase IV: Arrows indicate an arcade of post-flare loops following the eruption (panel (p)). (An animation of this figure is available.)}
\label{fig:EUV171_mosiac}
\end{figure*}

\begin{figure*}
\epsscale{0.75}
\plotone{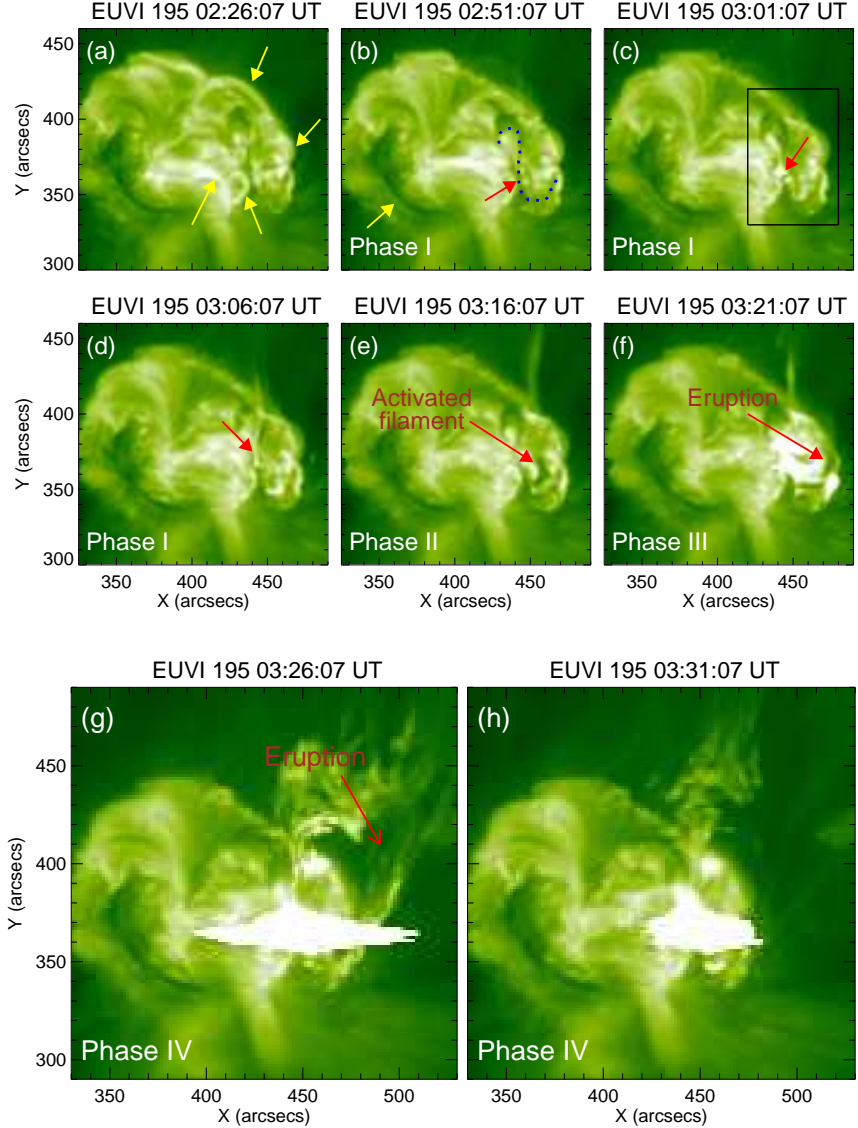}
\caption{Representative images taken by EUVI/SECCHI on board STEREO-B. The region of interest is shown inside the box in panel (c). In panel (b), we mark the prominence which underwent activation and subsequent eruption by a dotted line while another prominence belonging to this active region (shown by the yellow arrow) remains quiet throughout these observations. We note that the site of prominence activity is connected to other parts of active region by multiple loop systems (some of these loops are marked by arrows in panel (a)) which suggests that multiple flux systems close to the polarity inversion line are involved in the eruption process. In panels (d)--(g), we show a sequence of the dynamical activities occurring near the prominence during its pre-eruption and activation phases.} 
\label{fig:euvi_stereo}
\end{figure*}

\begin{figure*}
\epsscale{1.0}
\plotone{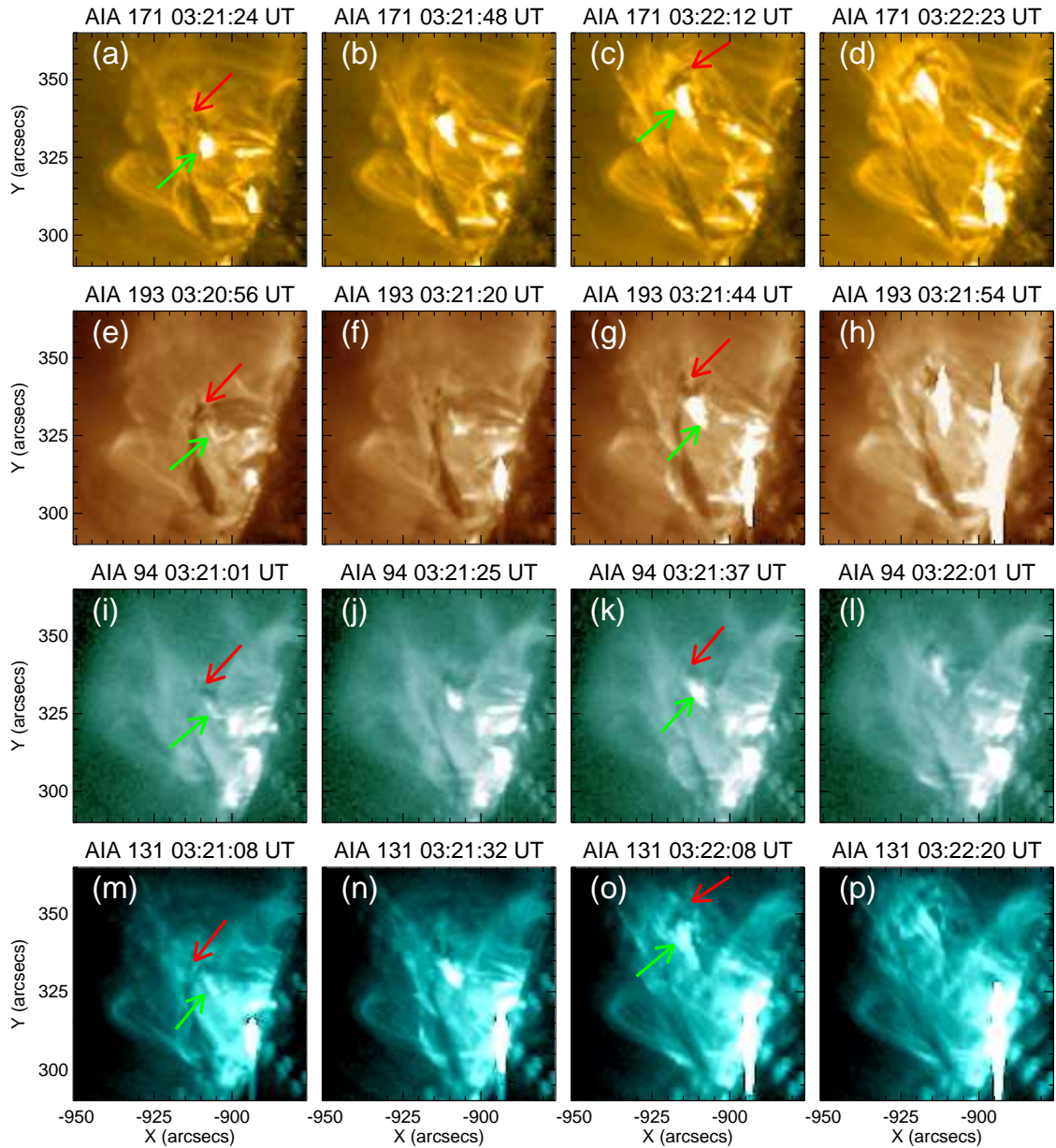}
\caption{Representative AIA images taken at different EUV channels showing the eruption of the prominence (i.e., fast expansion; phase III). In these images, we can clearly distinguish the fast eruption of the plasmoid (green arrows) below the fast rising prominence (red arrows). Note that the plasmoid was ejected into the direction of the apex of prominence.}
\label{fig:EUV_multi}
\end{figure*}

\begin{figure*}
\epsscale{1.0}
\plotone{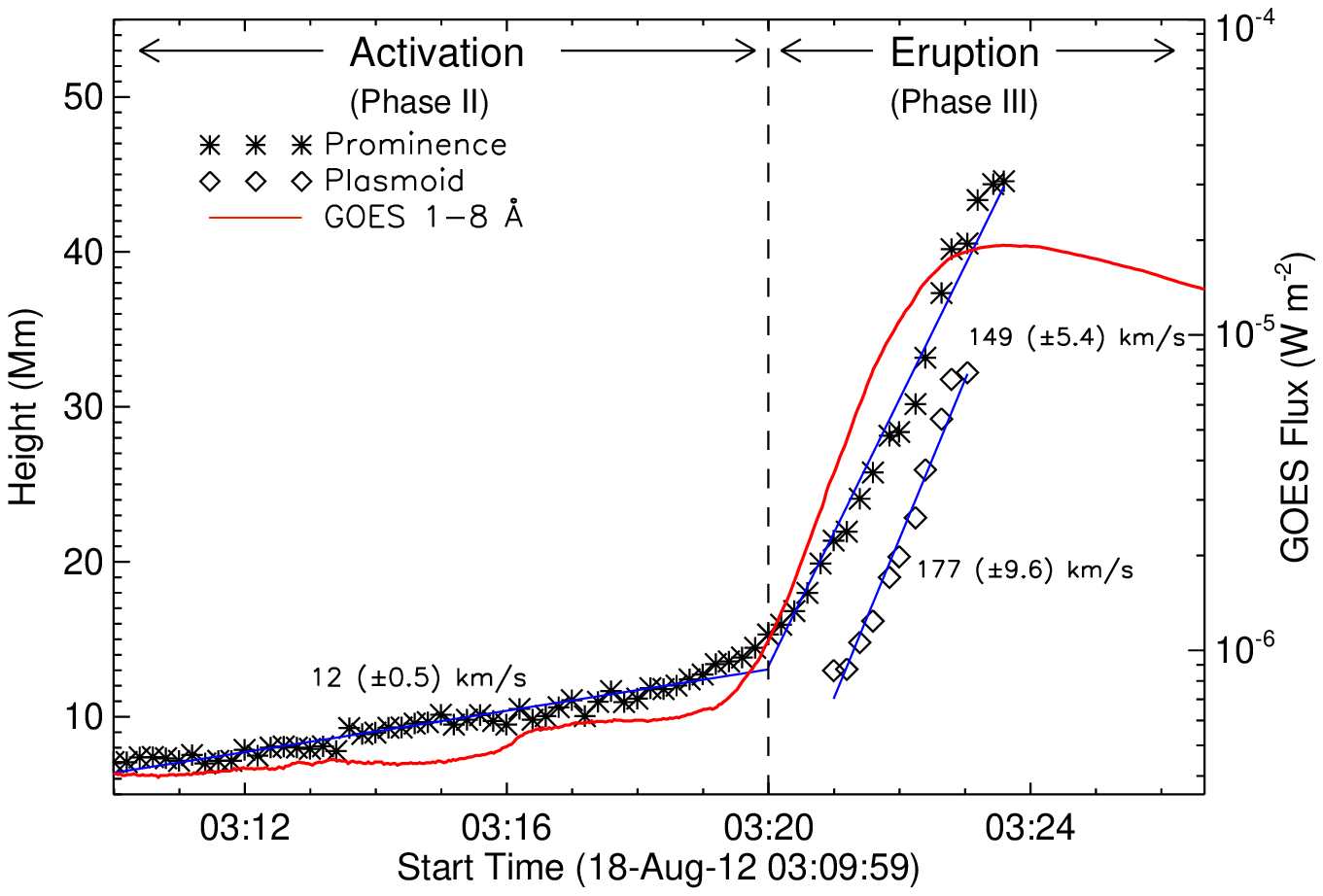}
\caption{Height-time plot of the prominence derived from AIA~171~\AA~images. After significant coronal activities during the pre-eruption period (phase I), the prominence activated and underwent a slow rise (03:10--03:20~UT; phase II) which is followed by the fast expansion during the eruptive phase (03:20--03:25 UT; phase III).  A few representative images during this period are presented in Figure~\ref{fig:EUV171_mosiac} where the prominence apex is marked by `$\times$' symbol. We have also shown the height-time plot for the plasmoid eruption (derived from AIA 171~\AA~images) which was observed below the apex of prominence during its eruptive phase. It is to be noted that the plasmoid was observed in other hot channels of AIA (Figure~\ref{fig:EUV_multi}) and underwent similar altitude evolution. The speeds of rising structures are estimated by a linear fit to the height-time data. The errors in speed estimations correspond to the standard deviation ($\sigma$) obtained from the respective linear fittings. Solid red line represents GOES 1--8~\AA~light curve.}      
\label{fig:prominence_h_t}
\end{figure*} 

The active region NOAA 11548 was at the east limb of the Sun on 2012 August 18. The examination of this active region between 02:45 and 03:45 UT displays a host of dynamical activities which are eventually connected to the activation and spectacular eruption of a prominence. By a careful investigation of the active region during this period along with light curves at different EUV and X-ray bands, we divide the whole activity in four phases which are summarized in Table~\ref{tab:phases}. In Figure \ref{fig:EUV171_mosiac}, we show some representative images captured by the 171~{\AA}~channel of AIA to depict the four evolutionary phases of the prominence eruption.  

The phase I (02:45--03:10 UT) corresponds to activities during the pre-eruption period (Figure~\ref{fig:EUV171_mosiac}(a)-(d)). The sequence of images clearly show that the corona above the activity site (i.e., the site of prominence rise) was quite dynamical during phase I. We particularly highlight two important developments during this phase. First, a collimated stream of plasma appeared at $\sim$02:47~UT, which extends higher into the corona (marked by arrow in Figure~\ref{fig:EUV171_mosiac}(a)). Such dynamic, transient and collimated coronal features are traditionally referred to as the jets \citep[see e.g.,][]{Moore2013}. EUV images clearly show that there is a continuous outward flow of hot plasma which leads to the upward extension of the spire of the jet.  
%Based on the morphological and kinematic evolution of this structure at multi-wavelengths, we recognize this as a blowout jet \citep{Moore2010,Moore2013}. We find that this blowout jet produces stronger emission and more dynamical evolution at cooler EUV lines (e.g., 304 \AA). 
Bright, well defined shape of the jet diminishes after $\sim$02:54 UT although such structures reoccurred later (03:04--03:05, 03:10--03:16~UT) at the same location.  
Cospatial to the jet activity, we note the striking emergence of a flux rope during 02:54--3:10~UT. This flux rope initially appeared to be a single, wide structure (marked in Figure~\ref{fig:EUV171_mosiac}(b)) which underwent a rapid upward expansion. The flux rope split into two parts as it grew in the corona while their lower portions are always connected (marked in Figure~\ref{fig:EUV171_mosiac}(c)). This flux rope appeared in absorption (i.e., dark structure) evidencing this to be at a lower temperature than the ambient corona. It is remarkable that the bifurcated structures continuously rotated (likely underwent untwisting) along their common base. With the upward, rotatory expansion of the flux rope, we observe plasma eruption along the spire of the previously observed jet and the widening of the spire. Due to the complex activities associated with the jet, we suggest this to be a non-standard jet and, following the nomenclature of \cite{Moore2010, Moore2013}, consider it under the blowout class of jets.

After about $\sim$30 minutes of dynamical activities in the corona (i.e., phase I), the prominence activation started and it underwent a slow rise during the phase II (03:10--03:20~UT; Figure 1(e)-(h)). The EUV images reveal that a brightening started below the rising prominence at $\sim$03:12 UT which continuously built up with the increase of prominence height. 
The prominence was set to fast rising (eruptive) motion during phase III (03:20--03:25~UT; Figure 1(i)-(l)). 
The source region, however, exhibited intense brightening for several minutes after the complete eruption of the prominence from the AIA field of view and a growing arcade of post-flare loops appeared (phase IV; Figure 1(m)-(p)). We also note that the eruption of this prominence was associated with a CME.

In Figure~\ref{fig:prominence_h_t}, we provide a height-time plot of the erupting prominence. We show the rise of the prominence during the activation (phase II) and eruptive (phase III) stages by placing a cross ($\times$) symbol at the apex of the prominence in different panels of Figure \ref{fig:EUV171_mosiac}.
A linear fit to the height-time data suggests that the prominence underwent a slow expansion with a speed of $\sim$12~km~s$^{-1}$ during the phase II. The prominence exhibited eruptive expansion during the phase III with a speed of  $\sim$149~km~s$^{-1}$. In Figure~\ref{fig:prominence_h_t}, we also present the GOES 1-8~\AA~ light curve to compare the dynamical evolution of the prominence with the brightening in the source region.

\subsection{Coronal structures of the active region}
After defining the phases of prominence evolution in the preceding section, we explore the morphology of active region loops and their environment with respect to the prominence and its activation. Since the active region is located just at the limb as per SDO observations, EUVI/SECCHI images on board STEREO-B enable us to have its on disk perspectives. In Figure~\ref{fig:euvi_stereo}, we provide a series of EUVI images at 195~{\AA} covering different phases of this activity.  The activity site is marked by a box in Figure~\ref{fig:euvi_stereo}(c). 

Although EUVI has a coarse temporal resolution (5 min), this data is extremely useful to understand the magnetic configuration of the activity site. We see that the activity site is connected to other parts of the active region by large bright loops. We mark four most noticeable loops in Figure~\ref{fig:euvi_stereo}(a) that have likely participated in the eruption of the prominence under investigation while the prominence is marked by dotted lines in Figure~\ref{fig:euvi_stereo}(b). The early pre-flare signatures can be seen in the form of localized brightening at the middle of the filament (marked by an arrow in Figure~\ref{fig:euvi_stereo}(c)). There is another filament channel in this active region (marked by yellow arrow in Figure~\ref{fig:euvi_stereo}(b) which does not take part in the eruption and remains undeflected throughout. The images show other important features, such as, pre-flare brightening (panel (c)), streams of hot plasma ejection (panel (d)), prominence activation along with source region brightenings (panels ((e)-(f)) and finally prominence eruption (panels (g)-(h)) which are well consistent with the observations taken at multiple EUV channels of AIA. 

The EUVI data confirms the occurrence of multiple dynamical activities in the vicinity of a twisted filament channel before its activation. The presence of large EUV coronal loops connecting the activity site to distant parts of the active region suggests that multiple flux systems existed close to the polarity inversion line (PIL) which is delineated by the prominence itself. 
 
\subsection{Filament's fast rise and plasmoid ejection}
\label{sec:plasmoid_ejection}

\begin{figure*}
\epsscale{1.0}
\plotone{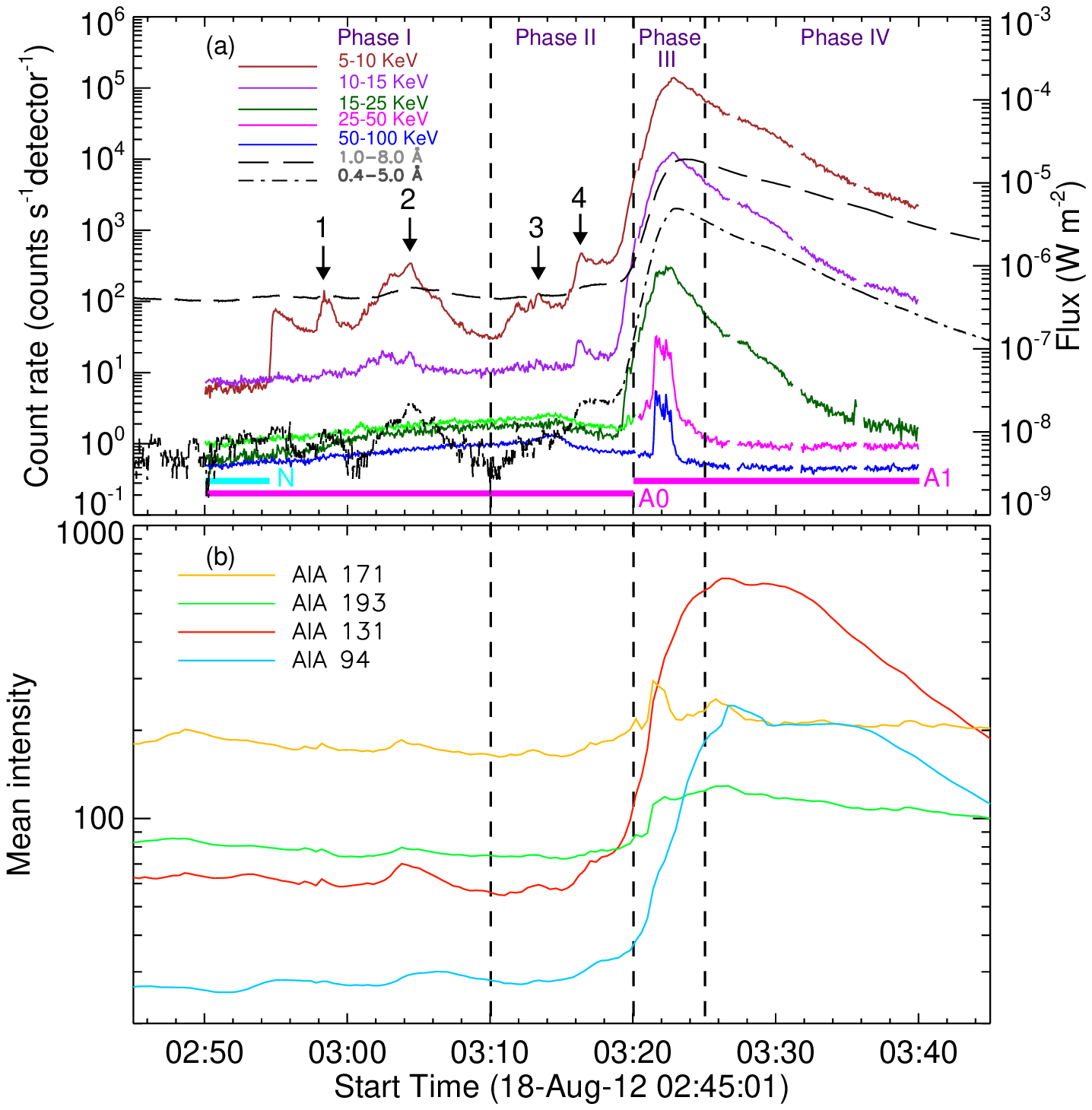}
\caption{Top panel: X-ray time profiles observed by GOES and RHESSI. The RHESSI light curves are presented for energy bands of 5--10, 10--15, 15--25, 25--50 and 50--100 keV with a time cadence of 4 s. The sudden increase in 5--10 keV X-ray flux at $\sim$02:54~UT arises due to the end of RHESSI night phase. Horizontal bars at the bottom represent RHESSI attenuator state (which changes from A0 to A1 at 03:20~UT) and RHESSI night (N). Various pre-flare peaks during phase I and II are marked as 1, 2, 3, and 4. Bottom panel: The EUV light curves at four wavelengths. These light curves denote the mean intensity of the activity sites derived from successive AIA images. For clarity of presentation, we have scaled RHESSI count rates by factors of 1, 1, 1/5, 1/10, and 1/30 for 5--10, 10--15, 15--25, 25--50 and 50--100~keV energy bands, respectively.}   
\label{fig:light_curve}
\end{figure*}

During the eruptive phase of the prominence (i.e., phase III), the images at several EUV wavelengths (that represent plasma at different temperatures) reveal very crucial features that required special attention. In Figure~\ref{fig:EUV_multi}, we present the multi-wavelength view of this phase at four channels of AIA, viz., 171~\AA~(log(T)=5.8), 193~\AA~(log(T)=6.2, 7.3), 94~\AA~(log(T)=6.8), and 131~\AA~(log(T)=5.6, 7.0). 

After the onset of fast eruption of the prominence at $\sim$03:20~UT (described in section~\ref{sec:prominence}), we observe intense brightening below the rising prominence which is different from the source region brightening (Figure~{\ref{fig:EUV171_mosiac}(i)). We notice detachment of a bright structure from the source region as the prominence further rose ($\sim$03:21 UT) and the structure continuously moved upward in the corona in consort with the prominence eruption (Figure~\ref{fig:EUV_multi}). This feature is seen at different AIA channels which reveals that it represents eruption of a blob of plasma at high temperature, i.e., a plasmoid. The rise of the prominence and the plasmoid is indicated in various panels of Figure~\ref{fig:EUV_multi} by red and green arrows, respectively. The plasmoid is always located below the rising prominence and clearly seen till $\sim$03:23~UT. In Figure~\ref{fig:prominence_h_t}, we plot the altitude evolution of this plasmoid from AIA~171~{\AA} images. We find that the plasmoid underwent a fast ejection with a speed of 177~km~s$^{-1}$.

\subsection{Episodic energy release and M1.8 eruptive flare}
\label{sec:eruptive}

\begin{figure*}
%\epsscale{0.7}
\plotone{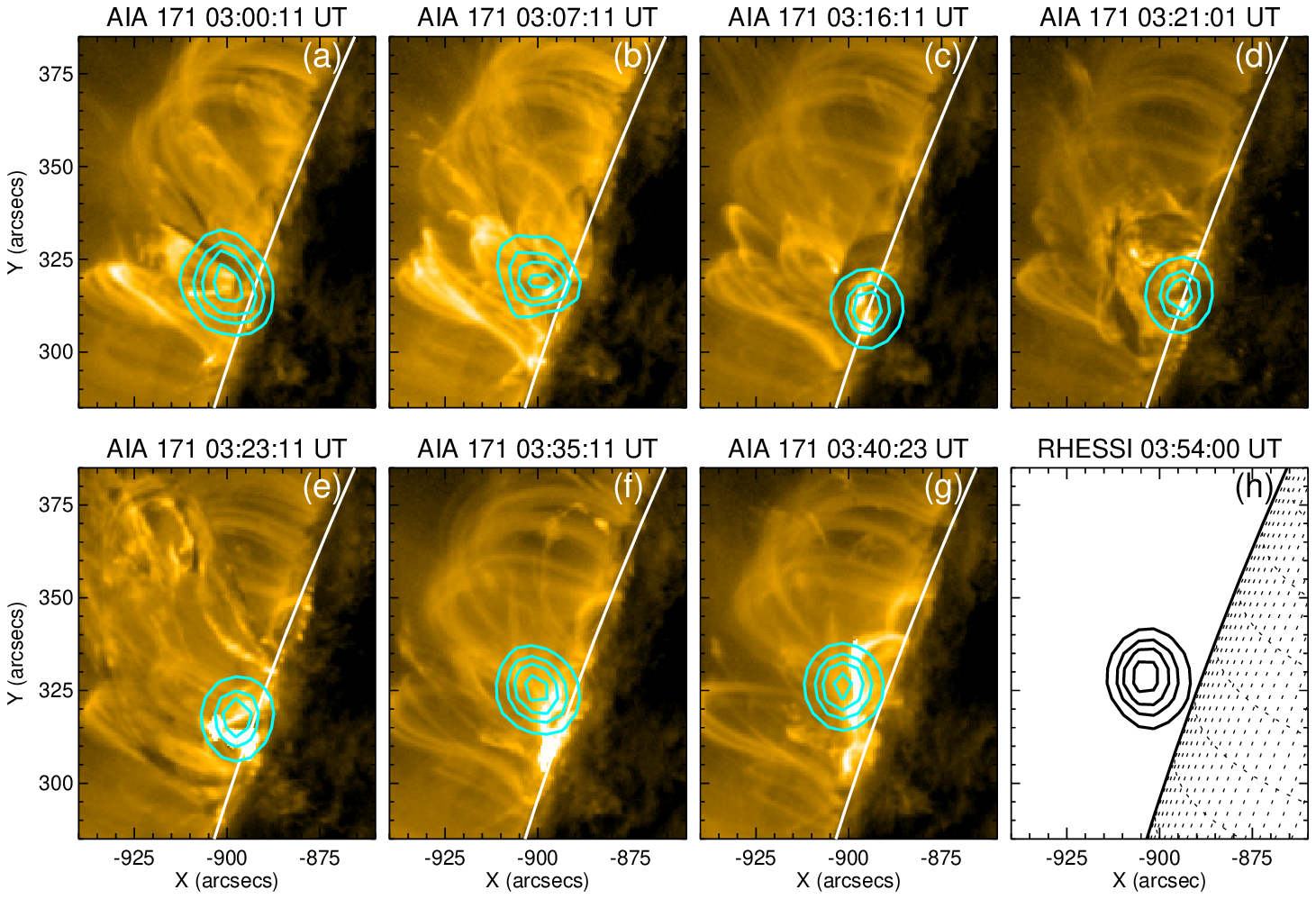}
\caption{A few representative images showing the evolution of RHESSI 6--12 keV source from the early pre-flare stages to the decay phase of M1.8 flare. To show the spatial association between the prominence evolution and the thermal X-ray source, AIA 171~\AA~images are plotted in the background (panels (a)--(g)). RHESSI images are reconstructed with CLEAN algorithm. Panels (a), (b), (f)--(h): The contour levels for X-ray image are 60\%, 75\%, 85\% and 95\% of the peak flux in each image. Panels (c)--(e): In order to compare the X-ray sources and corresponding EUV brightenings, the contour levels for X-ray images are set as  60\%, 80\% and 90\% of the peak flux in each image.}
\label{fig:LT_source_AIA211}
\end{figure*}

During the pre-eruption, activation, and eruptive phases of the prominence, we observe brightenings at different regions of the corona (see Figure \ref{fig:EUV171_mosiac} and section~\ref{sec:prominence}). These brightenings correspond to events of energy release at different temporal and spatial scales. In order to understand the temporal characteristics of the energy release phenomena, we present different light curves of the flaring region in Figure~\ref{fig:light_curve}. 

RHESSI observed the Sun continuously during the various phases of this prominence activity right from the pre-eruption phase till the end of the M-class flare. We note that the RHESSI measurements during this whole period are contaminated by X-rays from a particle event\footnote{http://hesperia.gsfc.nasa.gov/hessidata/metadata/2012/08/18/} (i.e., the RHESSI detectors were hit by high-energy particles trapped in the Earth's radiation belt). However, the particle rates are low and remain mostly constant throughout the observations. Further, the reconstruction of X-ray images at different energy band up to 200~keV~(Figure \ref{fig:HXR_evolution}) yields clear structures of X-ray sources which indicates that particle event (which mostly affects observations at higher energy bands) has not effected the RHESSI measurements much up to $\sim$200~keV.

The top panel of Figure~\ref{fig:light_curve} provides X-ray light curves obtained from GOES (1--8~and 0.5--4~\AA) and RHESSI (5--10, 10--15, 15--25, 25--50, and 50--100~keV).  We note that the RHESSI's attenuator state was A0 until 03:20~UT, i.e. RHESSI observed with its highest sensitivity at low energies, and then changed to A1, i.e. the thin attenuator was in place. In the bottom panel, we show EUV light curves in four channels (171, 193, 94, and 131~\AA). We also mark different evolutionary phases of the prominence (section~\ref{sec:prominence}; Table~\ref{tab:phases}) in this plot for a comparison.

RHESSI X-ray light curves (5--10 and 10--15~keV; Figure~\ref{fig:light_curve}(a)) clearly show that there is a significant emission at soft X-ray energies during the pre-eruption phase (phase I) from the active corona with distinct and episodic emission peaks (marked as 1 and 2).
% with the latter one exhibiting prolonged and stronger emission. 
The X-ray intensity further built up after the onset of prominence activation (phase II or precursor emission) and two prominent peaks can be identified at this phase (marked as 3 and 4) which are superimposed on the continuously rising flux level. The X-ray flux in all the energy bands rapidly rises and maximizes during phase III. By examining HXR light curves at energies $>$25~keV, we find distinct HXR burst during $\sim$03:21-03:24~UT, signifying the impulsive phase of M1.8 flare. During this time, the low-energy RHESSI emission is steeply increasing, attaining the maximum of the M1.8 flare at  $\sim$03:23~UT.
%Following peak 4, the X-ray flux in all the energy bands underwent rapid increase in emission  . The high energy light curves (>25~keV) clearly show  attaining the maximum of M1.8 flare at $\sim$03:22 UT. 
We further note that HXR bursts at high energy bands (i.e., 25--50 and 50--100 keV) are broad and structured with rapid intensity fluctuations.

A comparison between light curves and EUV images suggests that various peaks during the pre-eruption phase (phase I) indicate intermittent and localized episodes of coronal energy release. It is noteworthy that among X-ray light curves, pre-flare events are highly distinguishable in 5--10 keV RHESSI X-ray light curve and higher energy band of GOES (i.e 0.5--4~{\AA}). During the slow rise of the prominence (phase II), the region below the expanding prominence exhibits intense emission in the form of localized EUV brightening (Figures~\ref{fig:EUV171_mosiac}(e)-(h)) along with X-ray emission at 6--12~keV energies (Figure~\ref{fig:LT_source_AIA211}(c)). During this period, flare light curves at lower X-ray energies and EUV channels exhibit gradual yet steady rise in the intensity of emission suggesting precursor flare brightening.   
The eruptive phase of prominence (phase III) is characterized by impulsive increase in the intensity of emission (i.e., impulsive phase of M1.8 flare) implying the onset of large-scale magnetic reconnection. At X-ray bands, the impulsive phase is recognized up to $>$200 keV energies. After the complete expulsion of the prominence, the X-ray light curves indicate decay phase of the flare. However, the EUV light curves show a prolonged emission after the flare's impulsive phase due to the bright emission from the post-flare loop arcade.
 
\subsection{Spatial and spectral evolution of X-ray emission}
\label{sec:flare}

\begin{figure*}
\epsscale{0.8}
\plotone{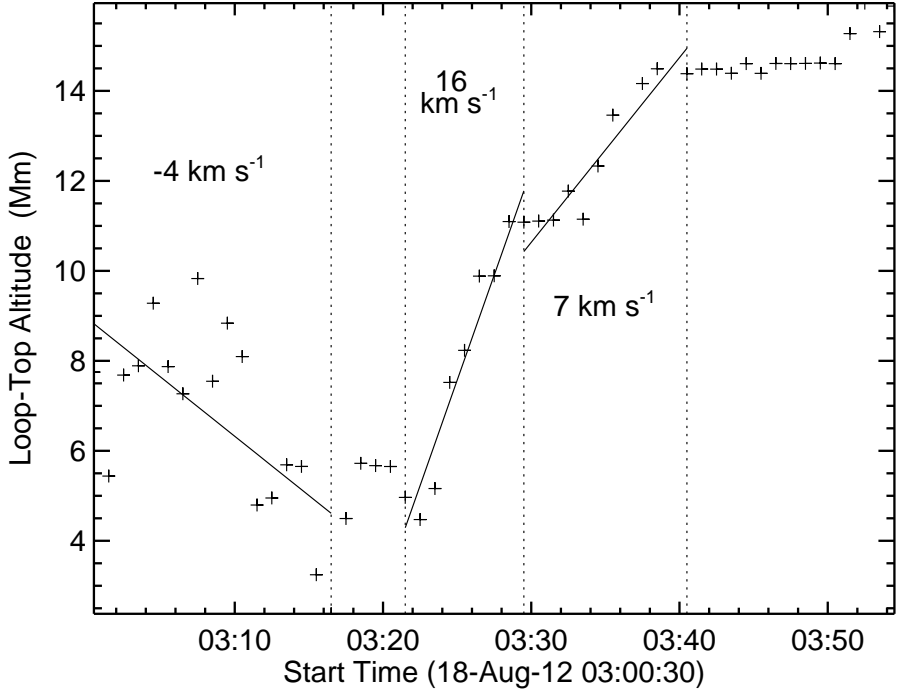}
\caption{Altitude evolution of the thermal 6--12~keV~source during phases of the prominence activation and eruption. We note that during the pre-flare phase (i.e.,$<$03:20 UT) X-ray source appeared at varying heights in the corona with a general trend of decrease in altitude. With the onset of the M1.8 flare at 03:20~UT, the X-ray source started exhibiting the `standard' upward motion.}
\label{fig:LT_evolution}
\end{figure*}

\begin{figure*}
\plotone{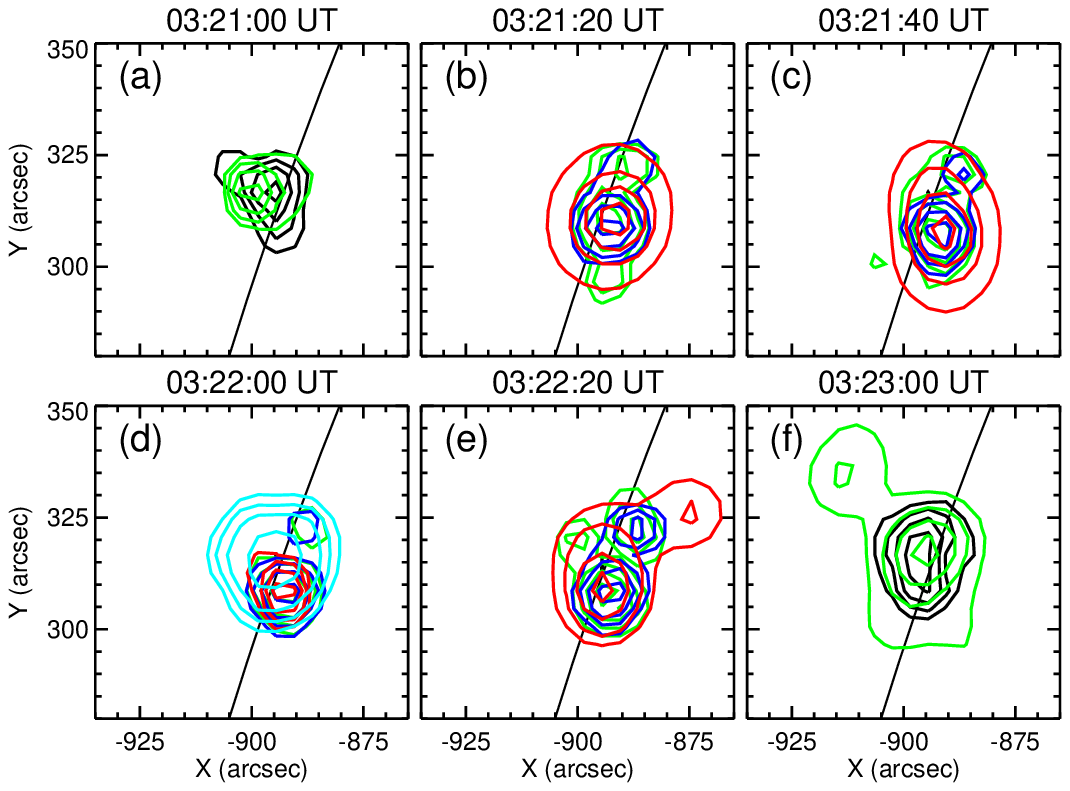}
\caption{Temporal evolution of HXR sources in  6--12 keV (sky), 12--25 keV (black), 25--50 keV (green), 50--100 keV (blue), and 100--200 keV (red) during the important stages of flare impulsive phase. These images are reconstructed with PIXON algorithm. The contour levels are set as 20\%, 35\%, 55\%, and 85\% of the peak flux of each images.\label{fig:HXR_evolution}}
\end{figure*}

The RHESSI images clearly reveal persistence of soft X-ray (SXR) sources (6--12~keV) during the whole period of the reported activity, i.e., from pre- to post- eruption phases (Figure~\ref{fig:LT_source_AIA211}).   
A careful examination of series of 6--12~keV~RHESSI images reveals that prior to the eruptive phase of the prominence, the SXR source appeared at varying heights in the corona with a general trend of decrease in altitude. The SXR source exhibits a steady rise after the onset of eruptive phase of the prominence (phase~III onward; $>$03:20 UT). In Figure~\ref{fig:LT_evolution}, we quantitatively present the altitude evolution of 6--12~keV~X-ray source from 03:00 to 03:54~UT. The source altitude is derived by measuring the centroids of X-ray emission along its main axis of motion. We find that, in general, altitude of the SXR source decreases during 03:00 UT to 03:16 UT with a speed of $\sim$4~km~s$^{-1}$. With the onset of eruptive phase, the source moved upward with a speed of $\sim$16~km~s$^{-1}$ till 03:30 UT which slowed down to $\sim$7~km~s$^{-1}$ thereafter.

The spatial evolution of X-ray sources at various energy bands, viz., 6--12 (sky), 12--25 (black), 25--50 keV (green), 50--100 (blue), and 100--200 keV (red), during the impulsive phase of M1.8 flare is shown in Figure \ref{fig:HXR_evolution}. We note a pair of distinct centroids at 50--100 keV and 100--200 keV energy bands that likely indicate emissions from conjugate footpoints of the flare loop systems (Figure~\ref{fig:HXR_evolution}(c)-(e)). The X-ray emission at low energies (6--12 and 12--25~keV) are observed in the form of single yet broad sources and indicate emissions from higher altitude (i.e., coronal emission; see Figure~\ref{fig:HXR_evolution}(a), (d), (f)). It is striking to note a HXR coronal source at 25--50 keV that underwent rapid altitude evolution (Figures \ref{fig:HXR_evolution}(e) and (f)). This HXR coronal source appeared for a brief interval ($ \sim $2 min) and evolved along the path of the erupted filament and plasmoid (section \ref{sec:plasmoid_ejection} and Figure \ref{fig:EUV_multi}). 
However, unlike to EUV plasmoid, it could not be followed to higher altitudes, presumably due to decrease in the density of corona.

In Figure~\ref{fig:sp_param}, we present background-subtracted temperature (T) and emission measure (EM) derived from GOES measurements. We note that the active region, which gave rise to the flare studied here, was also the origin of a long duration event of M5 class which peaked about two hours earlier and exhibited gradual decay phase for over an hour. Therefore, in order to chose a suitable background for T and EM estimations from GOES data, we selected a time interval between 02:50:00-02:51:00~UT during which the GOES flux was at the minimum level between the two events. We find a few intervals in the pre-eruption phase during which T and EM plots exhibit a flat profile (e.g., see T and EM plots around~03:10~UT). These flat sections indicate that the flux during the corresponding observing periods is comparable to the background flux. The plot also show T and EM derived from RHESSI spectroscopy for a few selected time intervals. For RHESSI analysis, we first generated a RHESSI spectrogram with an energy binning of 1/3 keV from 6 to 15 keV and 1 keV from 15 to 100 keV, and 5 keV from 100 to 200 keV energies. We only used front segments of the detectors, and excluded detectors 2 and 7. The spectra were deconvolved with the full detector response matrix. 
We have carefully chosen separate background intervals for different energy ranges: 02:56:30--02:57:30~UT for 6--15 keV and 03:09:30--03:10-30~UT for $>$15 keV. We believe that the choice of above mentioned background intervals would be useful to better distinguish the contributions from the background solar flux and particle event (non-solar event). As noted in Section~\ref{sec:eruptive}, the particle event was of low intensity and exhibited mild variations during the observing period (see section~\ref{sec:eruptive}). For reference, we present a few representative spatially integrated, background subtracted RHESSI spectra along with their respective residuals in Figure \ref{fig:hessi_spec}. Spectral fittings were obtained using the bremsstrahlung and line spectrum of an isothermal plasma and a non-thermal thick-target bremsstrahlung model. The spectra look quite reasonable within the observing limitations.

\section{Discussions}
\label{discuss}

\subsection{Pre-eruption activity and activation phase of the filament}

\begin{figure*}
\epsscale{0.7}
\plotone{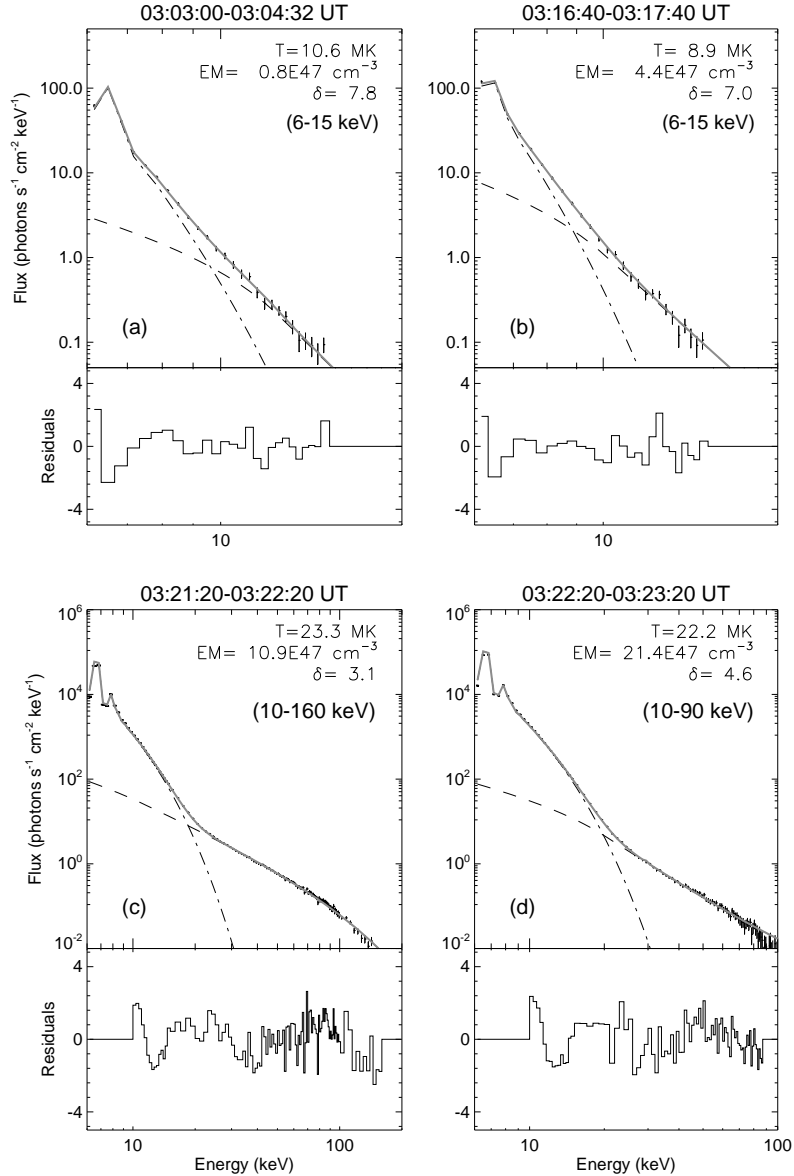}
\caption{RHESSI X-ray spectra derived during selected intervals along with their respective residuals. The spectra shown in panels (a) and (b) correspond to the interval prior to the eruptive phase of the prominence (i.e., phases I and II respectively). Panels (c) and (d) display spectra during the eruptive phase of the prominence and associated M1.8 flare (i.e., phase III). These spectra are fitted with a combination of an isothermal model (dashed-dotted line) and a thick-target bremsstrahlung model (dashed line) with the gray line indicating the sum of the two components. The energy ranges chosen for fitting are shown on each panels.}
\label{fig:hessi_spec}
\end{figure*}

The sequence of activities associated with the prominence eruption on 2012 August 18 clearly demonstrate that this eruption was associated with multiple evolutionary phases. These phases are preliminary defined in terms of the dynamical evolution of prominence while the X-ray and EUV light curves of the flare (and corresponding images) present crucial information about timescale,  intensity, and nature of energy release phenomena associated with various stages of the eruption. The energy release during a flare, as evidenced from a light curve, can exhibit diverse forms: small and slow (precursor phase), large and rapid (impulsive phase), or large and gradual (decay phase of large eruptive events). Notably, comparison of light curves at different energy levels provide valuable information about the timescales of energy release processes that operate simultaneously at different coronal and chromospheric heights.  

First we discuss the energy release events during the pre-eruption phase (phase I). The phase I is characterized by two pre-flare events (identified as 1 and 2 in Figure~\ref{fig:light_curve}(a)). The event 1 lasted for a shorter time period with smaller flux enhancement while event 2 exhibited prolonged and stronger emission with gradual rise and fall in the intensity of emission. 
%There are three events during this phase (identified as 1 and 2 in Figure~\ref{fig:light_curve}(a)). The third event is by far the largest and presents gradual rise and fall while the earlier two events lasted for a shorter time period with smaller flux enhancements. All these events present intermittent energy release processes with RHESSI light curves indicating a sequential increase in peak intensity for successive events. 
These peaks (i.e., events 1 and 2) are also identified in EUV light curves (Figure~\ref{fig:light_curve}(b)) although the enhancement in the flux level during the peak varies from one to another wavelength. The multi-wavelength data analyzed in the preceding sections clearly demonstrate that these energy release events of shorter timescales occurred in the vicinity of a pre-existing prominence that subsequently underwent an activation phase. More importantly, these intermittent events are associated with dynamical phenomena in the active region corona, namely, a blowout jet which has spatial and temporal association with the dramatic eruption of a cool flux rope (Figures \ref{fig:EUV171_mosiac}(a)-(d)). During the evolution of the flux rope, we observe localized brightening within the flux rope and at its legs as it undergoes rotation and upward expansion (Figure~\ref{fig:EUV171_mosiac}(d)). Coronal jets are believed to involve interchange magnetic reconnection, i.e., reconnection between closed and open flux regions. In active regions, the open field corresponds to field lines with one remote footpoint located very far from the locally closed domain \citep{Shimojo2000}. \cite{Sterling2015, Sterling2016} have clearly noted a casual relationship between minifilament eruptions and jet activity. In view of their studies, we suggest that the eruptive activities at relatively smaller-scales (such as the eruption of the cool flux rope) may not only cause jets but can further lead to the destabilization of a large-scale filament. In our study, the timing and location of the sequence of pre-eruption activities with respect to the active region prominence imply a probable connection between discrete events of energy release and subsequent activation of the prominence. In many earlier studies, discrete and localized X-ray brightenings have been reported that occur several minutes prior to the prominence activation and associated flares at sites closer to the prominence \citep{Chifor2007,Joshi2011,Joshi2013,Kushwaha2015}. Such small-scale energy release is often termed as pre-flare activity. These studies further suggest that the pre-flare activity represents early signatures of prominence destabilization. 

We note that during this pre-eruption phase, the X-ray emission at 6--12 keV was observed at varying altitudes with a general trend of decrease in altitude (Figures~\ref{fig:LT_source_AIA211} and \ref{fig:LT_evolution}). 
RHESSI imaging and spectroscopy confirms this pre-flare SXR emission to be predominantly from corona and thermal. More importantly, altitude variation of this 6--12~keV source is associated with a pre-flare event (marked as event~2 in Figure~\ref{fig:light_curve}) of phase I during which the X-ray emission exhibits a gradual rise and decline. In view of this temporal consistency, we speculate that the observed variations in X-ray emission centroids represent shift in the position of energy release site in the corona in relation to filament activation and overlying coronal activities. Certainly this kind of X-ray source evolution is not consistent with the canonical picture of solar flares in which the SXR sources, evidencing emissions from hot coronal loops, move successively upward in the corona. These observations reveal a complex magnetic configuration of pre-eruption corona where pre-flare energy release took place. The altitude decrease of X-ray looptop source during solar flares has been reported in earlier studies \citep{Sui2003, Sui2004, Joshi2007, Veronig2006,Joshi2009}. However, compared to the present case, the earlier observations indicate a more systematic downward motions of the X-ray looptop source. Further, in contrast to earlier observations, where X-ray sources underwent altitude decrease during the early impulsive phase of solar flares, we found altitude decrease of X-ray source during the modest energy release characterized by the pre-flare activity. In view of these differences, we speculate that the altitude decrease of X-ray source reported in this study is probably of different kind from the previously observed descending motions of X-ray sources.

%The filament activation starts with the onset of phase II. 
The filament is set to slow rise during phase II with a speed of 12~km~s$^{-1}$ (Figure~\ref{fig:prominence_h_t}). The X-ray light curves of this phase clearly reveal a continuous rise of the SXR  flux (see RHESSI 5--10 keV and GOES 1--8~\AA~profile; Figures~\ref{fig:light_curve}). This flux enhancement is superimposed by two peaks (marked as peaks 3 and 4) with peak 4 being noticeable at higher energies also (i.e., 10--15 keV profile). Although both phase I and phase II are associated with events of episodic energy release, the synthesis of multi-wavelength data suggests different characteristics of energy release phenomena. The phase I is characterized by varieties of coronal activity above the prominence and corresponding intermittent coronal EUV brightenings (Figure~\ref{fig:EUV171_mosiac}(a)-(d)). Further, it is noted that during phase I, RHESSI sources indicate X-ray emission from relatively higher altitudes (Figure~\ref{fig:LT_source_AIA211}(a)-(b)). On the other hand, once the prominence is set to upward expansion (i.e., onset of phase II), we do not notice significant coronal activities above the prominence and the X-ray and EUV brightenings are now observed continuously underneath the prominence (Figure~\ref{fig:LT_source_AIA211}(c)-(d)). 
%These observations imply that the gradual rise of SXR emission during phase II is likely the consequence of magnetic reconnection as predicted by the standard flare model although 
Notably, during this phase, the X-ray light curves reveal the energy release process to be small and gradual. Therefore, we interpret the energy release during phase II as a consequence of slow magnetic reconnection associated with prominence activation (i.e., its slow rising). 

By combining observations of phase I and II, we envisage the initiation of the eruption in the following steps. We recall that the episodic energy release during phase I is accompanied with a coronal jet along with other eruptive activities which essentially originate from the base of the jet. Under the framework of the canonical jet model, discussed earlier in this section, we argue that the processes during phase I would cause a weakening of the overlying magnetic fields that have restrained the high density prominence in the lower corona until this stage. It is noteworthy that the X-ray emission sources are confined and observed from the lower corona, close to the jet's base yet still above the apex of the prominence (Figure~\ref{fig:LT_source_AIA211}(a)--(b)). It has been suggested that coronal jets, manifesting reconnection opening of twisted fields, can drive solar eruptions \citep{Panesar2016}. We find that the X-ray intensity at the end of phase I reached the background level for the duration of about one minute which started to further build up with the slow rise of the prominence with the onset of phase II. Here is important to note that although the X-ray flux reveals a gradual decline for sometime between phase I and II, continuous mass motions are observed in the form of different eruptive activities in various EUV channels even during the stages of low X-ray intensity. Hence, we interpret that phase I and phase II are connected with the bulging of the envelop and core fields of the active region respectively while the X-ray emission is low, episodic, and localized.  

The on-disk view of the active region observed from EUVI/SECCHI clearly reveal that the filament under investigation was highly twisted `S-shaped' structure (Figure~\ref{fig:euvi_stereo}(b)). We clearly find that the filament is surrounded/overlaid by a complex system of coronal loops which together formed a multi-polar flux system in the corona. 
The occurrence of blowout jet as a prominent feature of the pre-eruption phase reveals that the multi-polar flux system of the active region contains open field lines. 
The jet activity is associated with the dynamical emergence of a cool flux rope which underwent rise and rotation simultaneously.   
Although it is hard to comment about the cause of this cool flux rope eruption, observationally it represents bulging of twisted magnetic field lines through the corona. It is very likely that the blowout jet along with cool flux rope eruption will cause magnetic reconnection at increasing coronal heights. Through the observations of these pre-eruption coronal activities, we argue that magnetic fields in active region corona, more importantly, the coronal region above the prominence, had sufficiently been weakened prior to onset of the prominence activation. With the activation of the prominence during phase II, we notice localized EUV brightening below the prominence with cospatial X-ray emission at energies $\lesssim$20~keV. At this time, the HXR emission is predominantly thermal with temperature upto $\sim$10~MK. The localized brightening at the lower corona and/or chromospheric heights during the slow rise of prominence essentially represents energy release within the core region. In view of these observations, we find that the radiative signatures during the slow rise phase of the prominence are in agreement with the {\it tether-cutting model} \citep{Moore1992}. 
Under this scenario, the rise of high density prominence would represent a situation when the constraining magnetic configuration had sufficiently weakened due to a combination of external (pre-eruption coronal activities) or internal (tether-cutting reconnection) processes. With the rise of the prominence, a vertical current sheet would form underneath it and reconnection in this current sheet would correspond to the flare brightening as proposed in the {\it standard flare model} \citep [i.e., classical ``CSHKP'' model of eruptive flares;][] {Carmichael1964,Sturrock1966,Hirayama1974,Kopp1976}.
%which is well consistent with our observations. 
In view of this, we recognize a multi-step process in the evolution of the prominence under investigation: the pre-eruption coronal activities are probably associated with the successive weakening of coronal magnetic fields in which prominence is embedded. The prominence activation is supported by the {\it tether-cutting} process while its steady growth is facilitated by an external process which ensures weakening of overlying fields.
% which will facilitate the outward expansion of core flux comprising the prominence. 
Following the steady rise of the prominence, the standard flare reconnection would set underneath which will further contribute toward the detachment of the prominence from the source region.

\subsection{Prominence eruption and M1.8 flare}

\begin{figure*}
\epsscale{0.8}
\plotone{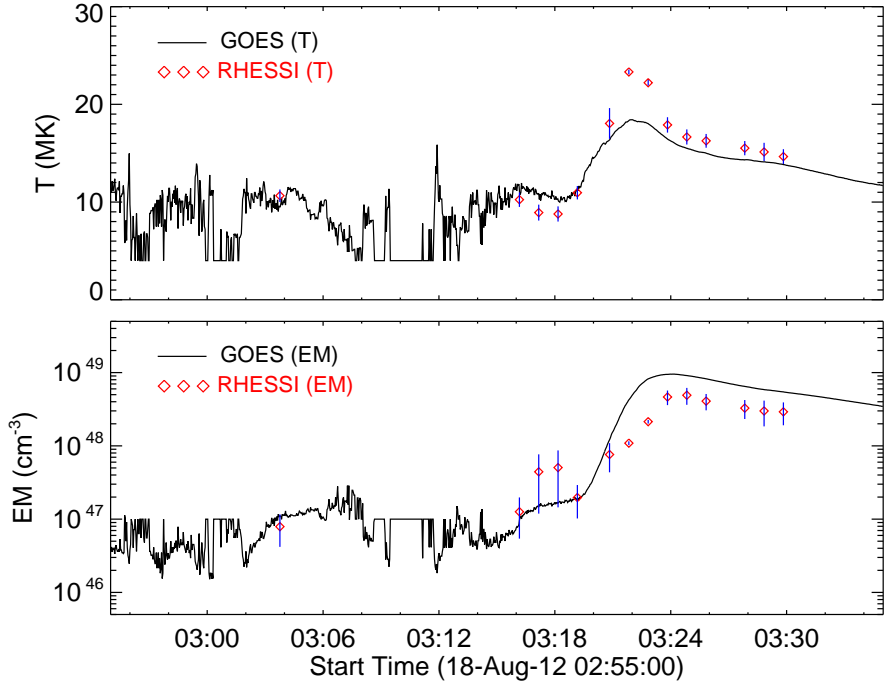}
\caption{Evolution of temperature (T) (panel (a)) and emission measure (EM) (panel (b)) derived from GOES and RHESSI X-ray measurements. The error bars are plotted on each T and EM values derived from RHESSI spectra.}
\label{fig:sp_param}
\end{figure*}

After significant pre-flare activities, the prominence was set to complete loss of equilibrium during phase III (Figures \ref{fig:prominence_h_t} and \ref{fig:light_curve}) and eventually erupted, triggering impulsive energy release in the source region that corresponds to the M1.8 flare. 
A comparison of the height-time plot of the erupting prominence with the GOES X-ray profile clearly reveals that the flare emission gradually built up during the prominence activation while it increased impulsively during the eruptive phase of the prominence (Figure~\ref{fig:prominence_h_t}). Such synchronization between the kinematics of the erupting flux rope and the energy release evolution in the associated flare has been reported in previous studies \citep{Zhang2001,Vrsnak2007,Temmer2008,Berkebile2012}. It is understood in terms of magnetic reconnection taking place in the  current sheet beneath the erupting flux rope that powers both the flux rope accelaration and the flare energy release. The RHESSI X-ray observations of the impulsive phase of the flare reveal high temperature (T $\sim$25 MK) and significant non-thermal characteristics ($\delta$ $\sim$3) indicating rapid plasma heating and acceleration of electrons to high energies up to 200 keV (Figures~\ref{fig:hessi_spec} and \ref{fig:sp_param}). This phase is accompanied with the upward motion of X-ray loop top source (Figure~\ref{fig:LT_evolution}). In the {\it standard flare model} this rising motion of X-ray source reflects the progression of magnetic reconnection during which field lines rooted successively further apart from the magnetic inversion line reconnect \citep[e.g.,][]{Joshi2007}. 

Intense brightening was observed from the source region right from the early stage of the prominence eruption. Soon afterward, the source region brightening became more structured and a distinct plasmoid appeared that underwent fast upward motion with the same direction of motion as the apex of the erupting prominence. The plasmoid eruption is observed in several EUV channels including 94~{\AA} and 131~{\AA} images that corresponds to very high plasma temperature ($\sim$10$^{6}$--10$^{7}$~K). In the unified model of solar flares \citep{Shibata1999}, plasmoid (or flux rope) ejection has been recognized as a key component of the flaring process, irrespective of the flare size. Evidences of plasmoid eruption in early observations were gathered from the flares observed from the Soft X-ray Telescope (SXT) on board Yohkoh \citep[see, e.g.,][]{Shibata1995, Ohyama1997, Ohyama1998,Nishizuka2010}. Recently evidences of plasmoid eruption was also found from EUV observations at different wavelength channels taken from AIA \citep{Takasao2012,Kumar2013}. From our observations, we note that the filament posed a large-scale structure comprised of dark material (i.e., cool and dense plasma) while the plasmoid (i.e., heated fluxrope) was observed as a blob of hot plasma below the apex of the prominence. It is noteworthy that during the impulsive phase of the flare the speed of the plasmoid ($ \sim $177 km s$ ^{-1} $) and the prominence eruption ($ \sim $149 km s$ ^{-1} $) are of the same order. By exploring the temporal and spatial correlations between the erupting prominence and plasmoid ejection, we recognize a feedback relation between the kinematic evolution of the prominence and large-scale magnetic reconnection \citep[see review by][]{Schmieder2015}. The fast eruption of the prominence would induce strong mass inflow into the current sheet underneath it which drives magnetic reconnection and particle acceleration. On the other hand, the ongoing magnetic reconnection would further reduce the downward acting tension of the overlying fields and provides additional poloidal flux to the flux rope, which is considered to be the main driver of the eruption \citep{Vrsnak2004, Temmer2008, Temmer2010}.  
%Thus, our observations imply that the onset of impulsive energy release (signifying fast reconnection below the outward moving prominence; cf. Figure~\ref{fig:prominence_h_t}) caused acceleration of plasmoid which subsequently supported the eruption of prominence in the lower corona.  

\section{Conclusions}

\begin{table*}
\begin{center}
\caption{Summary of various phases associated with the prominence activation and eruption. Note that various phases of the prominence eruption are marked in EUV and X-ray time profiles shown in Figure~\ref{fig:light_curve} and representative images recorded in the AIA 171~\AA~channel are presented in Figure~\ref{fig:EUV171_mosiac}.}
\begin{tabular}{p{2.0in}p{0.9in}p{3.1in}}
\tableline\tableline
Phases / characteristics  & Time (UT) & Observations\\
\tableline
Phase I: Pre-eruption coronal activity & 02:45 -- 03:10  & Dynamical activities in the corona: blowout jet associated with cool flux rope eruption,  intermittent energy release in thermal X-rays (5--15 keV) and EUV. Filament remained stationary during this phase.\\
 & &   \\

Phase II: Prominence activation & 03:10 -- 03:20 & Slow rise of filament with a speed of 12~km~s$^{-1}$, gradual increase of thermal X-ray and EUV emissions. EUV and X-ray brightenings are mostly confined underneath the prominence. \\
& &   \\

Phase III: Prominence eruption and M1.8 flare & 03:20 -- 03:25  & Prominence was set to fast rising motion with speed of 149~km~s$^{-1}$ and eventually erupted, multi-band EUV observations of plasmoid eruption below the erupting prominence, impulsive energy release in the source region leading to M1.8 flare.\\
& &   \\

Phase IV: Flare decay and post-flare loops & 03:25 -- 03:45 & Formation of post flare loops and arcades in the source region, slow decay of flare soft X-ray and EUV radiations. \\

\tableline
\label{tab:phases}
\end{tabular}
\end{center}
\end{table*}
    
\begin{itemize}

\item The pre-eruption phase reveal striking activities in the corona that extended upto larger coronal heights, much above the apex of the prominence. These activities include a blowout jet and dramatic evolution of a cool flux rope. 
%These activities occur in conjunction with intermittent X-ray and EUV brightenings at different locations/heights in corona. 
The intermittent rise in the intensity of the jet along with localized coronal brightenings associated with a cool flux rope eruption during phase I suggest that magnetic reconnection proceeded at various heights in the corona. In spite of these complex activities in the corona, the prominence remained in a quiet state at the core of this active region for several minutes before showing slow rise during phase II. Although it is unclear if the pre-eruption activities played a direct role toward triggering the prominence activation, we believe that these activities would certainly make a condition favorable for the prominence eruption by weakening the overlying magnetic fields that restrain the prominence.  
Thus we propose that a combination of external (i.e., pre-eruption coronal activities) and internal ({\it tether-cutting} reconnection) processes have contributed toward the successful eruption of the prominence.

\item The onset of the prominence activation (i.e., slow rise during phase II) is accompanied by a gradual increase of thermal X-ray and EUV emissions underneath it. The gradual rise of thermal emission corresponds to the precursor phase of the flare and suggests a more intense plasma preheating. We identify localized and gradual emission under the prominence as the slow magnetic reconnection resulting from {\it tether-cutting} process. The X-ray loop source continued to exhibit downward altitude shift until this phase. These observations reveal that the prominence destabilization is complex, multi-step process. From these observations, we conclude that the prominence activation is supported by the {\it tether-cutting} process while its steady growth is facilitated by an external process which ensures weakening of overlying fields.

\item After dynamical activities in the corona during which the prominence remained static and its subsequent slow activation, the prominence erupted resulting in the impulsive emission of M1.8 flare during phase III. The onset of violent eruption of the prominence indicates a catastrophic loss of equilibrium. Here we present a remarkable observation in which a plasmoid ejection immediately chased the erupting prominence. During this phase, the X-ray loop source (which exhibited downward motion in the two earlier phases) started moving upward, in consort with the magnetic reconnection scenario in the {\it standard flare model}. In view of these observations, we propose that the impulsive phase represents fast magnetic reconnection in a long vertical current sheet that probably lies between the plasmoid (in higher corona) and the X-ray coronal source (in lower corona). 
By examining the temporal, spatial and kinematic correlations between the erupting prominence and the plasmoid, we argue that the magnetic reconnection supported the fast ejection of prominence in the lower corona.

\end{itemize}

In this study, we aimed to address some important questions pertaining to triggering mechanism of solar eruptions and the process of prominence destabilization. Further, our study sheds some light about the role of magnetic reconnection towards the early evolution of erupting prominences. We propose to undertake more studies on these contemporary issues. 
\acknowledgments
We thank SDO, RHESSI, STEREO, and GOES teams for their open data policy. SDO and RHESSI are NASA's missons under Living With a Star (LWS) and SMall EXplorer (SMEX) programs, respectively. A.M.V. gtratefully acknowledges support from the Austrian Science Fund (FWF): P27292-N20. KSC is supported by the ``Planetary system research for space exploration" project of the Korea Astronomy and Space Science Institute. We sincerely thank the anonymous referee for providing
constructive comments and suggestions that have enhanced the presentation and quality of the paper.

%%%------------------------------
\bibliographystyle{apj}
\bibliography{New}
%%%%%%%----------------------

\end{document}